\documentclass[12pt]{iopart}
\usepackage{psfig}
\usepackage[dvips]{graphicx}
\begin{document}
\title{Bose-Einstein condensation of rubidium atoms in a triaxial
TOP-trap}

\author{J H M\"{u}ller,
D Ciampini, O Morsch, G Smirne, M Fazzi, P
Verkerk\footnote[3]{Permanent address: Laboratoire de Physique des
Lasers, Atomes et Mol\'{e}cules, Universit\'{e} de Lille 1, F-59
655 Villeneuve d'Ascq Cedex, France.},
 F Fuso, and E Arimondo}

\address{INFM, Dipartimento di Fisica, Universit\`{a} di Pisa, Via
Buonarroti 2, I-56127 Pisa, Italy}

\maketitle

\begin{abstract}
We report the results of experiments with Bose-Einstein
condensates of rubidium atoms in a triaxial TOP-trap, presenting
measurements of the condensate fraction and the free expansion of
a condensate released from the trap. The experimental apparatus
and the methods used to calibrate the magnetic trapping fields are
discussed in detail. Furthermore, we compare the performance of
our apparatus with other TOP-traps and discuss possible limiting
factors for the sizes of condensates achievable in such traps.
\end{abstract}




\section{Introduction}
Since the first observations of Bose-Einstein condensation (BEC)
in dilute alkali gases~\cite{wieman,hulet,ketterle}, experimental
as well as theoretical studies of degenerate quantum gases have
been published at an astonishing rate~\cite{ketterle2,stringari}.
Far beyond the mere realization and detection of BEC,
experimenters have investigated the static and dynamic properties
of Bose-Einstein condensates and have gained considerable control
over these macroscopic quantum objects, up to the point of
creating coherent beams of matter waves - atom lasers, in other
words. In spite of these early successes, experimental BEC is
still a growing and thriving field, and much research needs to be
done in order to test the vast number of theoretical predictions
made in the last few years.\\

In this paper, we present the experimental apparatus used to
create BECs of rubidium atoms in a triaxial
time-orbiting-potential (TOP) trap~\cite{toptrap}. To the best of
our knowledge, while the triaxial TOP trap has been used in BEC
experiments on sodium~\cite{hagley}, no previous application to
rubidium has been reported. We describe in some detail the
experimental parameters of our system and compare the performance
of our apparatus with those of other groups using similar setups.
Section 2 presents the experimental set-up, with emphasis on the
original parts for the rubidium cooling and transfer between the
two magneto-optical traps.  Section 3 reports the parameters for
the loading and evaporative cooling phases required to produce the
condensate. Moreover, the gain in phase-space density achieved
during the evaporation phases has been measured. In the following
sections the results of various measurements on the condensate are
reported. The final phase-space density, number of atoms and
temperatures associated to the different condensates are
presented. Furthermore, the expansion of the condensate following
a switch-off of the magnetic trap has been studied and compared to
different
 theoretical models. Finally, we describe different methods used for
precise measurements of the magnetic
 fields. In this way, we obtained an accurate calibration which was needed
as an input parameter for a
 theoretical model simulating the motion of the atomic
cloud~\cite{micromopaper}.

\section{Experimental setup}
\begin{figure}[h]
\centering{\psfig{figure=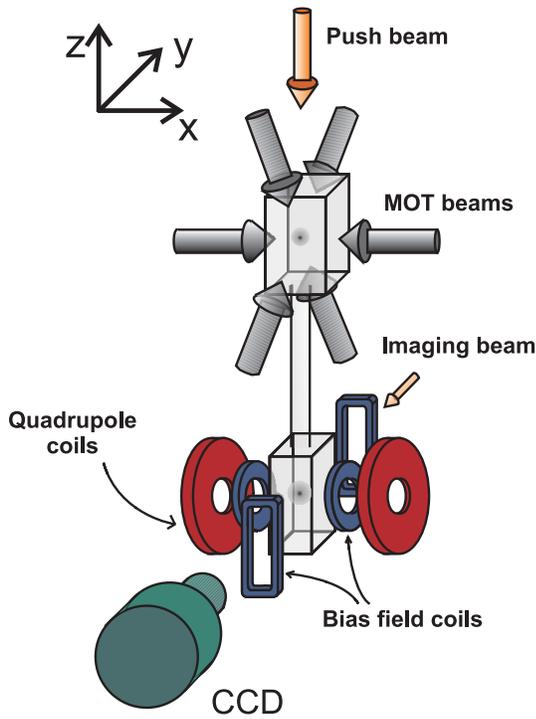,width=7.0cm}}
\caption[ ]{\label{setup}Setup of our experiment. The atoms are
transferred from the upper to the lower MOT by a push beam which
is briefly flashed on after loading the upper MOT for
$160\,\mathrm{ms}$. This cycle is repeated about $200$ times. For
clarity, the laser beams for the lower MOT are not shown here.}
\end{figure}

Our experimental apparatus is based on a double-MOT system with a
TOP-trap. The design of the vacuum system and the positioning of
the coils are shown in figure~\ref{setup}. Owing to the
arrangement of the quadrupole coils and the TOP-coils, our trap is
triaxial without cylindrical symmetry. In the following, we give a
brief overview of the specifications of our system.\\ {\em Vacuum
system:} Our vacuum system is composed of two quartz cells
connected by a glass tube of inner diameter $12\,\mathrm{mm}$ and
length $20\,\mathrm{cm}$ (see figure~\ref{setup}). At the upper
end of the glass tube, a graphite tube of length $6\,\mathrm{cm}$
and inner diameter $5\,\mathrm{mm}$ is inserted in order to
enhance differential pumping. The upper cell is connected to a
$20\,\mathrm{ls^{-1}}$ ion pump, whereas the lower cell is pumped
on by a $40\,\mathrm{ls^{-1}}$ ion pump in conjunction with a
Ti-sublimation pump. In this way, a pressure gradient is created
between the two cells with the pressure in the upper cell being of
the order of $10^{-8}\,\mathrm{Torr}$ and that of the lower cell
below $10^{-10}\,\mathrm{Torr}$. The upper cell also contains two
Rb dispensers (SAES getters) which we operate at
$3.0\,\mathrm{A}$.\\ {\em Lasers:} The laser light for the upper
and lower MOTs is derived from a MOPA (tapered amplifier) injected
in turn by a $50\,\mathrm{mW}$ diode laser. Under typical
conditions we extract up to $320\,\mathrm{mW}$ of useful output
from this system, which is then frequency-shifted by acousto-optic
modulators (AOMs) and mode-cleaned by optical fibres. In this way,
we create up to $60\,\mathrm{mW}$ of laser power for the upper MOT
and $15\,\mathrm{mW}$ for the lower MOT. The repumping light for
both the upper and the lower MOT is derived from a
$75\,\mathrm{mW}$ diode laser, yielding about $9\,\mathrm{mW}$ of
total power after passage through all the optical elements. The
injecting laser for the MOPA and the repumping laser are both
injected by $50\,\mathrm{mW}$ grating stabilized diode lasers
locked to Rb absorption lines.\\ {\em Magnetic trap:} Our TOP-trap
consists of a pair of quadrupole coils capable of producing field
gradients $2b^{\prime}$ (along the symmetry axis) in excess of
$1000\,\mathrm{Gcm^{-1}}$ for maximum currents of about
$230\,\mathrm{A}$, and two pairs of TOP-coils. The quadrupole
coils are water-cooled and are oriented horizontally (along the
$x$-axis, see fig.~\ref{setup}) about the lower glass cell of our
apparatus. A combination of IGBTs and varistors is used for fast
switching of the current provided by a programmable current source
(HP6882) whilst protecting the circuits from damage due to high
voltages induced during switch-off. In this way we are able to
switch off the quadrupole field within less than $50\,\mathrm{\mu
s}$ even for the largest field gradients. The rotating bias field
$B_{0}$ is created by two pairs of coils: One (circular) pair is
incorporated into the quadrupole coils, whilst the other
(rectangular) pair is mounted along the $y$-axis. Within the
adiabatic and harmonic approximations, for an atom with mass $m$
and magnetic moment $\mu$ this results
 in a triaxial time-orbiting potential $V_{TOP}$ given by
\begin{equation}
V_{TOP}=\frac{4
\pi^{2}m}{2}\left(\nu_{x}^{2}x^{2}+\nu_{y}^{2}y^{2}+\nu_{z}^{2}z^{2}\right)
\end{equation}
with the following frequencies along the three axes of the trap in
the ratio $2:1:\sqrt{2}$, as introduced in~\cite{hagley}:
\begin{eqnarray}
\nu_{x}&=&\frac{1}{2\pi} \sqrt{\frac{2\mu} {mB_{0}} } b^{\prime}\\
\nu_{y}&=&\frac{1}{2\pi}
\sqrt{\frac{\mu} {2mB_{0}} } b^{\prime}\\ \nu_{z}&=&\frac{1}{2\pi}\
\sqrt{\frac{\mu} {mB_{0}} }
b^{\prime}. \label{frequencies}
\end{eqnarray}
The anharmonic and gravitational effects neglected in this
approximation will be discussed in section 5. The TOP-coils in our
experiment can produce a bias field $B_{0}$ of up to
$30\,\mathrm{G}$ and are operated at a frequency of
$10\,\mathrm{kHz}$.\\ {\em Imaging:} Detection of the condensates
is done by shadow imaging using a near-resonant probe beam. The
absorptive shadow cast by the atoms is imaged onto a CCD-camera.
With a camera pixel size of $9\,\mathrm{\mu m}$ and a
magnification of about $1.2$, we achieve a resolution of just over
$7\,\mathrm{\mu m}$. Most of our measurements are made after a few
milliseconds of free fall of the released condensate, when typical
dimensions are of the order of $10-30\,\mathrm{\mu m}$.

\begin{figure}[h]
\centering{\psfig{figure=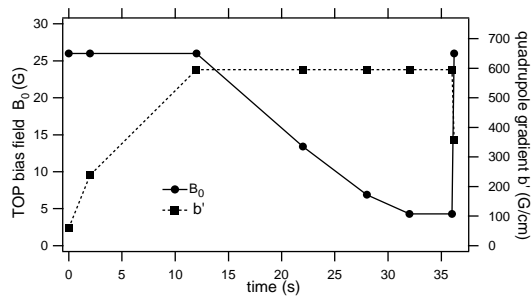,width=7.0cm}}
\caption[ ]{\label{ramps}Compression and evaporative cooling in
our TOP. Shown here are the ramps for the bias field $B_0$ and the
gradient along the $z$-axis, $b'$. The condensate is formed at t= 37
s, where both $B_{0}$ and $b^{\prime}$ are ramped for condensate
imaging.}
\end{figure}

\section{Evaporative cooling and creation of the condensate} 
\begin{figure}[h]
\centering{\psfig{figure=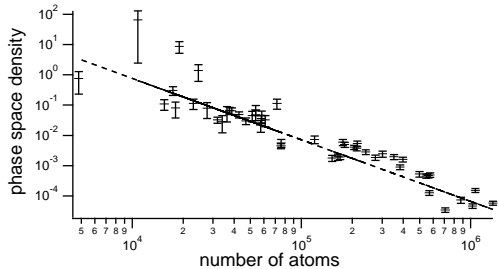,width=7.0cm}}
\caption[ ]{\label{phasedens}Phase-space density in the triaxial
TOP-trap as a function of atom number. The slope of the linear fit
in the log-log plot corresponds to a gain in phase-space density
of a factor 100 for a reduction in the atom number by a factor
10.}
\end{figure}

A typical
experimental cycle from the initial collection of atoms in the
upper MOT to the creation of a BEC is as follows. First, we load
about $5\times10^7$ $\mathrm{Rb}$ atoms into the lower MOT by
repeatedly (up to 200 times) loading the upper MOT for $\approx
160\,\mathrm{ms}$ and then flashing on a near-resonant push beam
that accelerates the atoms down the connecting tube. Once the
lower MOT has been filled, a $30\,\mathrm{ms}$ compressed-MOT
phase increases the density of the cloud, which is then cooled
further to about $15\,\mathrm{\mu K}$ by a molasses phase of a few
milliseconds. At this point, the molasses beams are switched off
and an optical pumping beam is flashed on five times for
$20\,\mathrm{\mu s}$, synchronized with the rotating bias field of
$1\,\mathrm{G}$ to define a quantization axis, in order to
transfer the atoms into the $|F=2,m_F=2\rangle$ Zeeman substate
desired for magnetic trapping. Transfer into the TOP-trap is then
effected by simultaneously switching on the rotating bias field
(at its maximum value of about $25\,\mathrm{G}$) and the
quadrupole field (at a value for the gradient chosen such as to
achieve mode-matching between the initial cloud of atoms and the
resulting magnetic trap frequencies). The subsequent evaporative
cooling ramps for the quadrupole and the bias fields are shown
schematically in figure~\ref{ramps}. After an adiabatic
compression phase, during which the quadrupole gradient is
increased to its maximum value, the bias field amplitude is ramped
down linearly. In this way, we perform circle-of-death evaporative
cooling down to a bias field of around $4\,\mathrm{G}$. Next, at a
constant bias field, we switch on a radio-frequency field,
scanning its frequency exponentially from $6.5\,\mathrm{MHz}$ down
to around $3.2\,\mathrm{MHz}$, which we find to be the threshold
for condensation for our system. At threshold, we have up to
$3\times10^4$ atoms in the condensate/thermal cloud-conglomerate.
Continuing rf-evaporation still further yields pure condensates of
up to $1-2\times10^4$ atoms with no discernible thermal fraction.
The value for the bias field at which we switch from
circle-of-death to rf-evaporation was chosen by maximizing the
final condensate number. The approach to BEC is illustrated
graphically in figure~\ref{phasedens}, in which the phase-space
density is plotted as a function of the number of atoms.\\ Before
imaging the condensate, we adiabatically change the trap frequency
by ramping the bias field and the quadrupole gradient in
$200\,\mathrm{ms}$. In this way, we can choose the frequency of
the trap in which we wish to study the condensate. Thereafter,
both fields are switched off on a timescale of $20-50\,\mathrm{\mu
s}$ for the quadrupole field and $100-200\,\mathrm{\mu s}$ for the
bias field. Owing to these short timescales, the change in trap
frequency during the switching can essentially be neglected as
typical oscillation periods in the trap are larger than
$10\,\mathrm{ms}$. In fact, we were able to observe non-adiabatic
motion of the trapped condensates at the frequency of the rotating
bias field~\cite{micromopaper}.

\section{Experimental results}
In the following, we briefly summarize some initial measurements
made on the condensates obtained with our apparatus.
\subsection{Evidence for condensation and condensate fraction}
\begin{figure}[h]
\centering{\psfig{figure=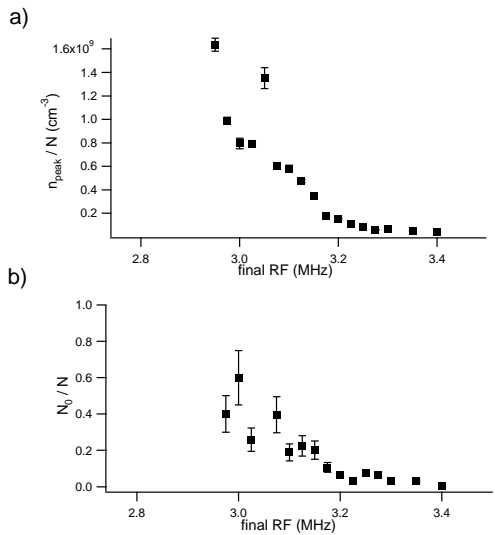,width=7.0cm}}
\caption[ ]{\label{condfrac}Normalized peak density of the trapped atomic
cloud (a) and condensate
fraction (b) as a function of the final RF-frequency. From the two graphs
it is evident that
condensation occurs at a final RF-frequency of about $3.2\,\mathrm{MHz}$.
The bottom of the trap is
located at a final frequency of $2.75\,\mathrm{MHz}$. Only data obtained
from absorption pictures
in which a non-condensed fraction was clearly discernible were included
in plot (b).}
\end{figure}

In order to find the threshold for condensation, the RF-frequency in the
final evaporation step is
lowered whilst monitoring the properties of the atom cloud (through shadow
imaging after
$3\,\mathrm{ms}$ of free expansion). At the threshold, the tell-tale signs
of condensation, namely
a sudden increase in peak density and the onset of a bimodal distribution,
begin to appear.
Figure~\ref{condfrac} shows plots of the peak density normalized with
respect to the number of
atoms (which removes the considerable experimental jitter especially in the
condensed regime) and
the condensate fraction as a function of the final RF-frequency. The
condensate fraction is
determined from a bimodal fit to single pixel rows of the absorption
picture, and it is evident in
the two plots that condensation sets in at a final frequency of about
$3.2\,\mathrm{MHz}$,
corresponding to a temperature of $365\,\mathrm{nK}$ as calculated from the
ballistic expansion of
the cloud, and a peak density of $\approx 5\times
10^{11}\,\mathrm{cm^{-3}}$. From this, we
calculate a phase-space density of $2.5$ at the threshold, in agreement
with theoretical
predictions. Using the expression $k_B
T_0=\hbar\bar{\omega}(N/\zeta(3))^{1/3}$ (valid in the
non-interacting approximation and with $\bar{\omega}$ equal to the
geometric mean of the three trap
frequencies) with $N=10^4$ atoms at the threshold~\cite{kasevich}, we find
$T_0\approx
400\,\mathrm{nK}$ in good agreement with our observed threshold
temperature.\\ We note here that,
unlike in the case of a static trap, for a TOP-trap there is no strict
proportionality between
$\nu_{cut}-\nu_0$ and $k_B T_{cut}$, where $\nu_{cut}$ is the frequency of
the RF-field, $\nu_0$ is
the resonance frequency at the bottom of the trap, and $T_{cut}$ is the
equivalent cut temperature.
A simple calculation considering the maximum instantaneous field at the
resonance shell shows that,
for low temperatures,
\begin{equation}
\nu_{cut}-\nu_0 = \frac{g_F }{h} \bigl(2k_B T_{cut}\mu
B_0\bigr)^{1/2}.
\end{equation}
This geometric average between the thermal cut energy  $k_B
T_{cut}$ and the magnetic energy in the bias field $\mu B_0$ of
the TOP-trap leads to a considerably more accurate control of the
cut energy in a TOP-trap. For instance, at a bias field of
$B_0=4\,\mathrm{G}$, a frequency difference $\nu_{cut}-\nu_0$ of
$350\,\mathrm{kHz}$ corresponds to a cut energy $T_{cut}$ of only
$1.2\,\mathrm{\mu K}$, whereas the same frequency difference in a
static trap leads to $T_{cut}=34\,\mathrm{\mu K}$.

\subsection{Free expansion of the condensate}
\begin{figure}[h]
\centering{\psfig{figure=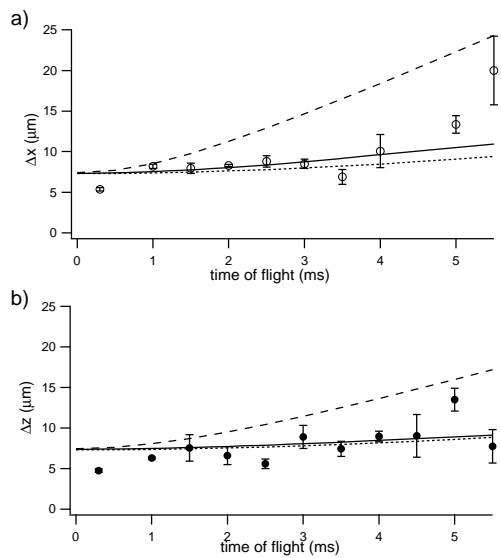,width=7.0cm}}
\caption[ ]{\label{expansion}Free expansion of a Bose-Einstein condensate
released from a triaxial
TOP-trap. The number of atoms in the (pure) condensates was around $10^3$.
Also shown are the
theoretical predictions for a harmonic oscillator wave-packet (dotted
line), the Thomas-Fermi limit
(dashed line), and a numerical integration of the full Gross-Pitaevskii
equation (solid
line). The theoretical graphs are corrected for a finite resolution of
$7\,\mathrm{\mu m}$.}
\end{figure}

One way of obtaining information on the properties of a Bose-Einstein
condensate is to investigate
its behaviour after it is released from the trap. Its subsequent evolution
is then monitored by
taking absorption images after a variable time-of-flight. The results of
such measurements on a
condensate released from a trap with $\nu_z = 363\,\mathrm{Hz}$ are shown in
figure~\ref{expansion}. Theoretically, the expansion of a condensate has
been investigated by
several authors, and analytical expressions for the condensate width and
its aspect ratio as a
function of time can be found in special cases. Figure~\ref{expansion}
shows the predictions of a
model based on the Thomas-Fermi approximation~\cite{castin+dum}, in which
the energy of the
condensate is dominated by the mean-field interaction between the atoms, as
well as the theoretical
expansion of a ground-state harmonic oscillator wavefunction, for which
interactions are neglected
entirely. Clearly, our experimental data agree with neither of these two
extremes. This is to be
expected, as the sizes of our condensates, with typically a few thousand
atoms in a pure
condensate, are rather small and therefore do not fully satisfy the
conditions for a Thomas-Fermi
treatment. It is, therefore, necessary to compare our data with a numerical
integration of the full
Gross-Pitaevskii equation. The results of such an integration are also
plotted in
figure~\ref{expansion}. As expected, they lie between the two extreme
models and fit our data
reasonably well. It is clear, however, that our condensate number is so low
that the interaction
term in the Gross-Pitaevskii equation is almost negligible and the
numerical results are close to
the pure harmonic oscillator case.

\section{Calibration of the magnetic fields}
In many applications of magnetic traps, it is sufficient to
describe the trap by its characteristic frequencies for dipolar
oscillations of atomic clouds. In such a measurement, one applies
a magnetic field along a chosen axis for a short time, thus giving
a kick to the (initially stationary) atomic cloud, and monitors
the subsequent oscillations of the atoms. With a judicious choice
of the points in time at which the position of the cloud is
sampled, one can achieve frequency measurements with uncertainties
well below the percent level. Deducing absolute values of the
magnetic field gradient and the bias field from these measurements
with similar accuracy, however, is not so straightforward. The
main incentive for us to accurately measure these absolute values
was that we needed them as input parameters for numerical
simulations of non-adiabatic motion in the
TOP-trap~\cite{micromopaper}. In the following, we shall briefly
describe several methods we used to measure absolute values for
both the quadrupole gradient and the bias field and indicate the
uncertainties associated with these measurements. For the most
part, the measurements were carried out with condensates, which
facilitated the determination of the position of the atomic
cloud.\\ In the first method, we measure the vibrational
frequencies $\tilde{\nu_x}$ and $\tilde{\nu_z}$ along the $x$- and
$z$-axes, respectively, exciting the dipolar modes along these two
directions simultaneously. In order to be able to use theoretical
formulas derived in the harmonic approximation taking into account
the effect of gravity, we have calculated the anharmonic
corrections up to fourth order, including cross-terms, following
the scheme presented by Ensher~\cite{ensher}. The  results
reported in Appendix A allow us to deduce from our measured
frequencies the corresponding values in the harmonic limit
(equivalent to infinitesimal oscillation amplitudes; typical
amplitudes in our experiment are between $20\,\mathrm{\mu m}$ and
$60\,\mathrm{\mu m}$.). Those anharmonic corrections can be up to
$1\%$ of the measured values and are, therefore, essential if an
accuracy in the magnetic field below the percent level is desired.
The quadrupole gradient can be calculated directly from the ratio
$\tilde{\nu_x}/{\tilde\nu_z}$ given by in the harmonic
approximation with the gravitational corrections by
\begin{equation}
\frac{\tilde{\nu_x}}{\tilde{\nu_z}}=\sqrt{2}\sqrt{\frac{1+\eta^2}{1-\eta^2}}
\end{equation}
 Here, $\eta$, defined by
\begin{equation}
\eta=\frac{\mu b'}{mg}
\end{equation}
measures the ratio of magnetic and gravitational forces along the
$z$-axis. It is interesting to note that in the triaxial TOP the gravity
corrections are equal to those derived for a cylindrically
symmetric TOP-trap \cite{ensher}. Re-substitution of the value for $b'$
thus retrieved
along with either of the two frequencies into the expression for
$\tilde{\nu_x}$ or $\tilde{\nu_z}$ then yields a value for $B_0$.
For instance, $\tilde{\nu_z}$ is given by
\begin{equation}
\tilde{\nu_z}=\frac{1}{2\pi}\sqrt{\frac{\mu}{mB_{0}}}b'\left(1-\eta^2\right)^{3/
4}.
\end{equation}
In order to check the obtained values for $B_0$ and $b'$ by
independent methods not relying on the calculated frequencies for
a TOP-trap, we use two separate strategies. In one method, the
quadrupole gradient is measured by first trapping and
evaporatively cooling atoms in the presence of both the quadrupole
and the bias fields. Then, the bias field is switched off, which
shifts the centre of the quadrupole potential with respect to the
TOP-potential. The quadrupole gradient is subsequently determined
by measuring the acceleration of the atoms and subtracting the
acceleration due to gravity. In this way, $b'$ can be determined
with a relative error of less than $1\%$. An independent
measurement of the bias field $B_0$ is made by switching off the
quadrupole field after the atoms have been cooled in the TOP
whilst leaving the bias field on. A short ($100-500\,\mathrm{\mu
s}$) RF-pulse is then applied to the atoms at a given frequency,
and the number of atoms remaining in the original trapped state is
measured after turning the quadrupole field back on (about
$1\,\mathrm{ms}$ after switching it off). When the frequency of
the RF-pulse matches the Zeeman-splitting due to the bias field,
atoms are transferred into untrapped Zeeman-substates and hence
lost from the trap. Using this method, we found two different
values of the RF-pulses for which atoms were lost from the trap,
indicating that there is a slight asymmetry between the magnetic
fields produced by the two pairs of TOP-coils. Measuring $B_0$
with this method proved to be less reliable than with the method
described above, but yielded the same value for the bias field to
within $5\%$.\\

\section{Condensate numbers in TOP-traps} 
\begin{figure}[h] \centering
\includegraphics[width=0.8\textwidth]{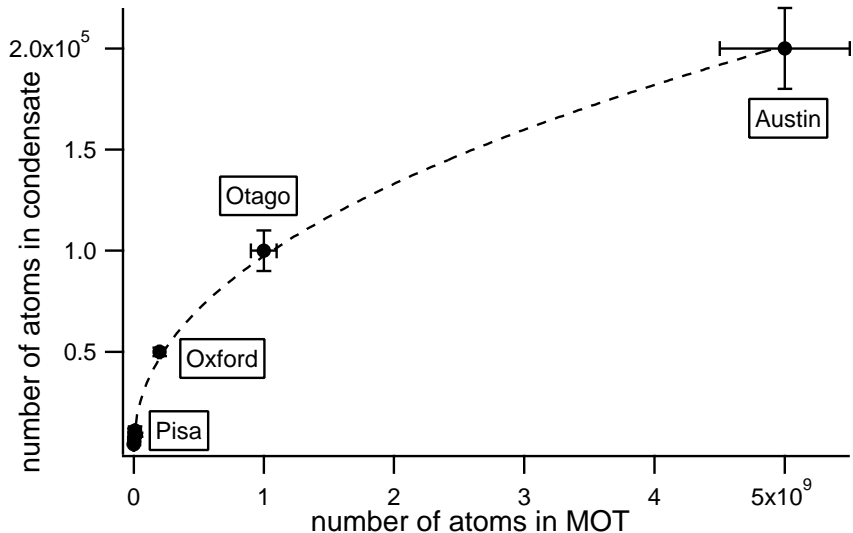}
\caption[ ]{\label{loading}Typical condensate numbers of various
groups as a function of initial MOT numbers. The data were taken
from publications of the groups at Austin~\cite{han},
Otago~\cite{otago}, and Oxford~\cite{oxford}.}
\end{figure}

In our experimental apparatus, we
obtain condensates containing up to a few $10^4$ atoms, starting
from typical MOT numbers of about $5\times10^7$. Extrapolating
this linearly, one would expect to achieve condensate numbers of
up to $10^6$ for an initial number of $5\times10^9$ atoms in the
MOT. In the literature, however, one typically finds reports of
some $10^5$ atoms in the condensate under such circumstances. In
figure~\ref{loading} we have plotted typical figures for the MOT
and the condensate numbers for a few groups using rubidium
TOP-traps. Evidently, the reported condensate numbers do not scale
linearly with the MOT numbers. Instead, they can be fitted roughly
by a square-root law. Varying the MOT numbers in our own
experiment, we find a similar behaviour on a smaller scale. We
discovered this when trying to increase the size of our
condensates and found that the main limiting factor comes from the
compression phase after loading the magnetic trap. Above a certain
number of atoms loaded into the MOT, we saw next to no increase in
the atom number after compression (or, for that matter, in the
condensate) when increasing the initial number of atoms. As
in~\cite{han}, we attribute this to an unfavourable ratio of the
size of the initial cloud and the circle-of-death radius. When the
cloud becomes too big, the circle-of-death cuts into it during
compression and thus any increase in the atom number is eaten up
by this cutting. This may be a limiting mechanism for most groups
and could explain the law of diminishing returns that is evident
in figure~\ref{loading}. In this context it is interesting to note
that, for instance, the JILA group uses a much higher bias field
($50\,\mathrm{G}$) than most other groups and achieves a much
better transfer efficiency from the MOT to the
condensate~\cite{cornell}, obtaining condensates of $\approx 10^6$
atoms for initial numbers of the order of $2\times 10^8$. Although
this may suggest that a larger bias field is the answer, it is not
clear whether there are other effects that limit the transfer
efficiencies achievable in TOP-traps.

\section{Conclusion}
We have presented the results of preliminary measurements on
Bose-Einstein condensates of rubidium atoms obtained in a triaxial
TOP-trap. Our experimental data for the condensation threshold and
the free expansion of the condensate agree well with theoretical
predictions. Increasing the number of atoms in our condensates
will allow us to further improve on the quality of our data and
investigate the properties of the condensates in more detail.

\section*{Acknowledgments}
O.M. gratefully acknowledges financial support from the European
Union (TMR Contract-Nr. ERBFMRXCT960002). This work was supported
by the INFM 'Progetto di Ricerca Avanzata' and  by the CNR
'Progetto Integrato'. The participation of G. Memoli and D.
Wilkowski in the early stages of the experiment is gratefully
acknowledged. The authors are grateful to R. Mannella for the
numerical integration of the Gross-Pitaveskii equation and to M.
Anderlini for help in the calculation of the anharmonic
corrections.

\begin{appendix}
\section{Anharmonic corrections in the TOP-trap}
For our calibration measurements, we deduced the frequencies in
the harmonic limit from the anharmonic corrections in the triaxial
TOP-trap. Terms containing the amplitude of the oscillations along
the $y$-axis have not been calculated as we do not excite
oscillations along that direction, but can be obtained in the same
manner. The expressions for the frequencies along the axis $i$
($i=x,z$) are then
\begin{equation}
\nu_i^{\rm anh}=\nu_{i}+\Delta\nu_i\quad ; \quad
\Delta\nu_i=\left(\frac{b'}{B_0}\right)^2\sum_j\alpha_{ij}a^2_j
\end{equation}
where $\nu_{i}$ is the frequency in the harmonic approximation, as
given by Eqs. \ref{frequencies}  and $a_j$ is the amplitude of the
oscillation in the $j$-direction. The elements $\alpha_{ij}$ of
the anharmonic correction matrix are given by
\begin{eqnarray}
\alpha_{xx}=\frac{\nu_{x}}{4}\left[6\frac{1-\eta^2}{1+\eta^2}\left(2-3\eta^2
-\frac{15}{8}(1-\eta^2)^2\right)-\frac{\eta^2(1-3\eta^2)^2}{18(1+\eta^2)}\right]
\\
\alpha_{xz}=\frac{\nu_{x}}{4}\left[\frac{7-\eta^2(18-15\eta^2)}{12}-\frac{9
\eta^2(3\eta^2-1)(14+8\eta^2)-8\eta^2}{36(7+9\eta^2)}\right]
\\
\alpha_{zx}=\frac{\nu_{z}}{2}\left[\frac{1-3\eta^2}{2}-\frac{2\eta^2(1-3\eta
^2)}{9(3+5\eta^2)}+\frac{7-3\eta^2-15\eta^2(1-\eta^2)}{12}\right]
\\
\alpha_{zz}=\frac{\nu_{z}}{2}\left[-\frac{3}{8}(1-\eta^2)(1-5\eta^2)-\frac{1
5}{8}\eta^2(1-\eta^2)\right].
\end{eqnarray}
\end{appendix}

\end{document}